\documentstyle[11pt, axodraw]{article}

\renewcommand{\b}{{\bf b}}
\newcommand{\q}{{\bf q}}
\newcommand{\k}{{\bf k}}

\begin{document}
\baselineskip=0.25in

\begin{titlepage}
\begin{flushright}
UM-TH-97-20 \\
hep-ph/9709497 
\end{flushright}

\bigskip

\begin{center}
{\Large {\bf Power Corrections and the Gaussian Form of the \\ 
 Meson Wave Function}}\\

\bigskip

{\bf R. Akhoury, \  A. Sinkovics and M. G. Sotiropoulos}\\

{\it The Randall Laboratory of Physics\\
 University of Michigan\\
 Ann Arbor, MI 48109-1120}\\

\end{center}

\bigskip

\begin{abstract}
\begin{quotation}
The wave function of a light pseudoscalar meson is considered and
nonperturbative corrections as signaled by perturbation theory 
are calculated. Two schemes are used, the massive gluon and the 
running coupling scheme. Both indicate the presence of leading power 
corrections of ${\cal O}(b^2)$, whose exponentiation leads to a Gaussian 
dependence of the wave function on the impact parameter $b$. The dependence
of this correction on the light cone energy fractions of the quark and 
the antiquark is discussed and compared with other models for the meson. 
\end{quotation}
\end{abstract}

\medskip

PACS 25.80.D, 13.40.G

\end{titlepage}

\newpage

\section{Introduction }
\setcounter{equation}{0}

The challenging task of explaining experimental data for exclusive processes, 
such as pion and nucleon electromagnetic form factors and elastic scattering, 
is tightly connected with our understanding of the bound states of light 
hadrons. 
In the asymptotic limit of parametrically large momentum transfer ($Q$) 
there is a clear theoretical picture that comes from merging the parton model 
with perturbative QCD (pQCD). The basic tenet of this picture is that a
hadronic bound state is a superposition of virtual states with a definite 
number of constituent partons. In the frame where all participating hadrons 
are fast moving, factorization guarantees that the $Q^2$ dependence of 
an exclusive observable enters only through the underlying process of 
the elastic scattering of the partonic constituents. This has lead to the,
by now classic, dimensional (or quark) counting rules \cite{dimcount}. 
However, the fact that the scaling with $Q^2$ as predicted by the counting 
rules is actually observed by experiments at high momentum transfers 
generated new questions. Counting rules assume that the dominant 
configurations are those in which none of the constituents carries vanishingly 
small longitudinal momentum. 
But such configurations are present in the hadronic wave function and 
would lead to scaling with $Q^2$ that is less steeply falling as 
$Q^2 \rightarrow \infty$. 
Underestimating the contributions from these end-point regions was the main 
criticism against the parton model picture \cite{criticism}. 

The next major step forward came with the work of Sterman and collaborators
\cite{BottsSter, SterLi} who implemented the summation of leading and next to 
leading  logarithmic radiative corrections through the introduction of 
Sudakov factors into the factorized expressions for exclusive processes. 
The generic form of such factors is 
$\exp\left[-c \ln Q^2  \, \ln (\ln Q^2/ \ln (1/b^2))\right]$ 
where $b$ is the transverse distance between constituents. 
It has the effect of suppressing configurations of constituents that are 
separated by large distance $b$, which would be the case in the end-point 
regions mentioned previously. 
We note that the Sudakov factors can be considered as the perturbative tail 
of the hadronic wave function, i.e. the region where the bound state 
properties can be reconstructed purely from pQCD.  
Sudakov improved perturbation theory, implemented with a model for the 
hadron wave function at low momentum scale
and a reasonable prescription on how to freeze the 
coupling at large scales, does give numerically stable results with the 
smallest possible number of phenomenological parameters 
\cite{SterLi, Li, BottsSterMGS}. 
This approach sets the benchmark but it is not the final answer. 
The main problem now becomes to estimate the  sensitivity 
to soft contributions of the Sudakov improved perturbation theory for 
moderate $Q^2$, that is  in the range 
\mbox{$3 {\rm GeV}^2 \le Q^2 \le 40 {\rm Gev}^2$}. 
To this end we need a framework for studying nonperturbative corrections 
to hadronic wave functions. This is the objective of this paper. 

Since our approach to the problem is from the high $Q^2$ end, we consider 
the two quark component of a light pseudoscalar meson wave function. 
This is the dominant configuration in the asymptotic limit. 
The specific exclusive process in which the pion participates will be of 
no concern here.
Our goal is to calculate the structure of the nonperturbative corrections 
as they are signaled by perturbation theory itself. 
These will occur in the form of power corrections in $b$ in the Fourier 
transformed wave function. 

Recently, some progress has been made in understanding the power corrections by
considering classes of Feynman diagrams that give rise to a factorial
divergence of the perturbation series in large orders. 
For the case of QCD the power corrections arise from the small momentum 
region of the loop integrations and are associated with the so called 
infrared renormalons. 
For a review and related references, see \cite{AkhZak, Beneke}. 
Thus, one may get an indication of the type of power corrections by looking 
at the infrared sensitivity of Feynman diagrams. 
Most investigations involving renormalons are in the context of two 
calculational schemes. 
The first is the massive gluon and the second is the running coupling scheme. 
In the first method, which we apply to the case of the radiative corrections 
to the meson wave function, a gluon mass is introduced via a dispersive 
parameter $\lambda$ \cite{dispersive} that acts as an  infrared regulator. 
This procedure is certainly consistent at the one loop level. 
Since the power corrections, in the renormalon approach arise from the 
infrared sensitive regions, our interest will be in the structure of the 
nonanalytic terms in $\lambda^2$ which, to the order we work in, 
turn out to be $\ln \lambda^2$ and $b^2 \lambda^2 \ln\lambda^2$. 
It is known \cite{AkhZak} that such nonanalytic terms come only from the 
pinch singular points in the loop momentum integration  with $\ln \lambda^2$
being associated  with the usual logarithmic enhancements of the perturbative
series and $b^2 \lambda^2 \ln \lambda^2$ with the leading power corrections.  
In the second scheme, power corrections are computed using a
one loop improved perturbation theory where infrared effects are introduced 
through the running coupling $\alpha_s(k_\perp^2)$.  
Then the Sudakov exponent is Borel transformed  and its singularity structure 
in the Borel plane is studied.  
The power corrections in this case are found to be proportional to 
$\Lambda_{\rm QCD}^2 b^2$. 
This technique is the same one used for the study of power corrections to 
inclusive processes and has been widely discussed in recent years,
see for example Ref.~\cite{KorchSter}.

In section 2 we calculate the leading power corrections 
and discuss the correspondence between the expressions derived in the  
two schemes.  
Exponentiation of  leading power corrections naturally lead to the 
Gaussian dependence of the meson wave function on the impact parameter $b$.
In section 3 we discuss our results  and compare them with previous 
publications where such Gaussian factors for the meson wave function 
have been advocated. 
Here, of interest will be the $x$ and $\ln Q^2$ dependence of the 
Gaussian factor itself, where $x$ is the light cone fraction of the 
momentum of the quark in the meson.
Finally we summarize our conclusions in the last section.

\section{Radiative corrections to the meson wave function}
\setcounter{equation}{0}

\subsection{Definitions}

The Bethe-Salpeter two quark wave function of a fast moving pseudoscalar 
meson $M$ is defined by the following matrix element at renormalization 
(and factorization) scale $\mu$ \cite{BottsSter}:  
\begin{equation} 
X(k,p,\mu) = \frac{1}{N_c} \, \int \frac{d^4 y}{(2 \pi)^4} 
\, e^{i k\cdot y} \, 
\langle 0 | T \left( \bar{q}(0) \bar{\not{v}}  \gamma_5 q(y) \right)
  | M(p) \rangle  \, .
\label{Xdef}   
\end{equation}
In the frame where the meson is moving fast we define the light cone 
vector along the direction of motion $v^\mu=(1/\sqrt{2})(1,0,0,1)$ and the 
parity reflected direction vector $\bar{v}^\mu=(1/\sqrt{2})(1,0,0,-1)$,  
$\mu=0,1,2,3$, normalized as $v \cdot \bar{v} =1$.
Then the light cone momenta of the meson and the quark constituent 
are defined as
\begin{equation}
p^\mu=p^+ \delta^\mu_+, \  \  k^+ = \bar{v} \cdot k =x p^+ ,  \ \ 
k^- = v \cdot k, \  \ k_{\perp} = \k . 
\label{lcdefs}
\end{equation}
The large scale is $p^+=Q/\sqrt{2}$, where $Q$ is the momentum transfer of the 
hard exclusive process in which the meson participates. 
Although we do not need to specify the process, we must always keep in mind 
that our discussion will always be in the context of a hard process. 
The object of interest in this work is the wave function in impact parameter 
space given by 
\begin{eqnarray}
{\cal P}(x, b, p, \mu) &=& \int d^2 \k \, e^{i \k \cdot \b} \, 
\int dk^- X(k, p, \mu) 
\nonumber \\
&=& \frac{1}{N_c} \, \int \frac{d y^-}{2 \pi} e^{i x p^+ y^-} \, 
\langle 0 | T \left( \bar{q}(0) \not{\bar{v}} \gamma_5 q(0, y^-, \b) 
            \right) | M(p) \rangle \, .
\label{Pdef}
\end{eqnarray}
The quark distribution amplitude, which enters the leading order 
perturbative expressions, is  
\begin{equation}
\phi(x, \mu) = \int^{|\k|=\mu} d^2 \k \int d k^- \, X(k,p,\mu) \, 
= {\cal P}^{(0)}(x, b=0, p, \mu) \, .
\label{distampl}
\end{equation}
Radiative corrections to ${\cal P}^{(0)}$ coming from the infrared region 
exponentiate. 
Their summation leads to the Sudakov suppression factor \cite{Collins}
\begin{equation}
{\cal P}(x, b, p,\mu) = \exp \left( -S_{wf}(x, b, Q, \mu)\right) \, 
{\cal P}^{(0)}(x, b=0, p, \mu) \, ,
\label{exponentiate}
\end{equation} 
with 
\begin{equation}
S_{wf}= \frac{2 C_F}{\beta_0} \ln \frac{Q^2}{\Lambda_{\rm QCD}^2} 
\ln \left( \frac{\ln Q^2/ \Lambda_{\rm QCD}^2}
{\ln 1/(b^2 \Lambda_{\rm QCD}^2)} \right)  + {\rm NL} \, ,  \   \   \   \ 
\beta_0= \frac{11}{3} N_c - \frac{2}{3} N_f \, ,
\label{sexplog}
\end{equation}
where NL denotes next to leading logarithmic corrections, which are also 
known. 
Note that the above perturbative answer is really defined in the region  
$1/b^2 \gg \Lambda_{\rm QCD}^2$, i. e. the evolution of the wave function 
can be reliably constructed only for small size bound states.

The Sudakov exponent shown in Eq.~(\ref{sexplog}) is obtained from the 
one loop radiative corrections to ${\cal P}^{(0)}$. 
There exist various methods of calculating these corrections. 
In this paper we will mostly follow the approach of Botts and Sterman 
\cite {BottsSter}. 
In this approach the fermions are taken massless and the axial gauge is used. 
The advantage of using the axial gauge in the derivation of an evolution 
equation for the wave function is that the diagrams involved in the 
calculation of the kernel have simpler structure than in a covariant gauge. 
Specifically, diagrams in which gluons couple the kernel with the 
'inside' of the wave function thus generating non-2PI connections in the 
constituent quark channel (non-ladder gluons) are suppressed, Fig.~1.
In the covariant gauge such diagrams must be included and further reduced 
through the use of Ward identities, just like in the treatment of 
parton distribution functions in D.I.S.

\medskip

\begin{center} 
\begin{picture}(150, 80)(0,0)
\Curve{(0, 58) (25, 58)}
\Curve{(0, 62) (25, 62)}
\ArrowArcn(60, 20)(60, 130, 42)
\ArrowArcn(60,100)(60, 318, 230) 
\BCirc(20,60){8}
\Oval(60,60)(10,8)(0)
\Gluon(28, 60)(52, 60){1}{9}
\Gluon(60, 70)(58, 80){1}{3}    
\Gluon(60, 50)(58, 40){1}{3}
\Text(20,60)[]{$X$}
\Text(5,70)[]{$p$}
\Text(74, 60)[]{$q$}
\Text(40, 67)[]{$l$}
\LongArrow(78, 63)(78, 53)
\Text(105,61)[]{$\times$}
\Text(110,60)[l]{$\frac{1}{N_c} \bar{\not{v}} \gamma_5 e^{i {\bf b \cdot q}}$}
\end{picture} 
\end{center}
\vspace{-1.5cm}
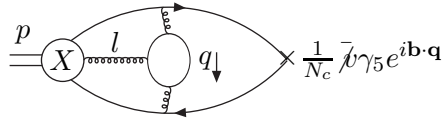
{\bf Fig.1} Class of diagrams that contain 'non-ladder' gluons ($l$). 
These are suppressed in the axial gauge.

\medskip
 
In the axial gauge there are gauge ($n$) dependent contributions to 
$S_{wf}$ that enter in the  next to leading logarithmic corrections in 
Eq.~(\ref{sexplog}). 
Such gauge dependent pieces cancel against gauge dependent terms in the 
radiative corrections to the hard scattering. 
Since we do not analyze the hard subprocess here, our results will be 
at the level of leading logarithmic and power corrections.

\subsection{Meson wave function in the massive gluon scheme}

As was mentioned in the introduction, we are interested in the 
nonanalytic dependence of the radiative corrections to the meson 
wave function on the infrared cutoff.  
The  Landau equations guarantee that infrared sensitivity 
arises from those momentum regions where the internal lines in a Feynman graph 
approach their mass shell.
To regularize mass-shell divergences we  introduce for the gluon a 
(dispersive) mass parameter $\lambda$ \cite{dispersive} and calculate the 
one loop radiative corrections to ${\cal P}^{(0)}$ in this massive gluon 
scheme. 
This is the main difference with the treament found in Ref.~\cite{BottsSter} 
and also the fact that we are mainly interested in the terms 
that are nonanalytic in $\lambda$ and vanish as powers of $\lambda$ in the 
exact mass-shell limit.
The gluon polarization tensor $N_{\mu \nu}$ is defined via the principal 
value prescription for the unphysical singularities $n\cdot q$. 
Hence the gluon propagator reads: 
\begin{eqnarray}
D_{\mu \nu}(q, n) &=&\frac{i}{q^2-\lambda^2+i \epsilon} N_{\mu \nu}(q, n) \, , 
\nonumber \\
N_{\mu \nu}(q,n)&=& 
 - g_{\mu \nu} +\frac{ n \cdot q}{(n \cdot q)^2+ \eta^2} \, 
n_{\left( \mu \right. } q_{\left. \nu \right)} 
- \frac{n^2}{(n \cdot q)^2+\eta^2} \, q_\mu q_\nu  \, ,
\label{propdef}
\end{eqnarray}
with $\eta \rightarrow 0$. 
At the end of this subsection we comment on the possibility of using the 
same parameter $\lambda$ to regularize the unphysical $n\cdot q$ 
singularities. 
The graphs to be calculated are depicted  in Fig.~2 and the Dirac structure 
of the meson-$q$-$\bar{q}$ vertex is 
\mbox{$ {\bf X} = -(1/4) \not{v} \gamma_5 X $}, where $X$ is the scalar 
wave function defined in Eq.~(\ref{Xdef}).

\medskip

\begin{center} 
\begin{picture}(350, 120)(0,0)


\Curve{(0, 58) (25, 58)}
\Curve{(0, 62) (25, 62)}
\ArrowArcn(60, 20)(60, 130, 42)
\ArrowArcn(60,100)(60, 318, 230) 
\BCirc(20,60){8}
\GlueArc(60, 120)(55, 300, 240){1}{20}
\Text(20,60)[]{$X$}
\Text(5,70)[]{$p$}
\Text(22, 76)[]{$k$}
\Text(60, 60)[]{$q$}
\LongArrow(64, 60)(71, 60)
\Text(105,60)[]{$\times$}
\Text(60, 10)[]{($a_1$)}

\Curve{(110, 58) (135, 58)}
\Curve{(110, 62) (135, 62)}
\ArrowArcn(170, 20)(60, 130, 42)
\ArrowArcn(170,100)(60, 318, 230) 
\BCirc(130,60){8}
\GlueArc(170, 0)(55, 58, 120){-1}{20}
\Text(130,60)[]{$X$}
\Text(170, 60)[]{$q$}
\LongArrow(166, 60)(159,60)
\Text(215,60)[]{$\times$}
\Text(170, 10)[]{($a_2$)}


\Curve{(240, 58) (265, 58)}
\Curve{(240, 62) (265, 62)}
\ArrowArcn(300, 20)(60, 130, 42)
\ArrowArcn(300,100)(60, 318, 230) 
\BCirc(260,60){8}
\Gluon(291, 79)(291, 41){1}{10}
\Text(260,60)[]{$X$}
\Text(245,70)[]{$p$}
\Text(266, 78)[]{$k$}
\Text(298, 60)[]{$q$}
\LongArrow(298, 54)(298, 47)
\Text(346,61)[]{$\times$}
\Text(350,60)[l]{$\frac{1}{N_c} \bar{\not{v}} \gamma_5 e^{i {\bf b \cdot q}}$}
\Text(300, 10)[]{($b$)}
\end{picture} 
\end{center}
\vspace{-0.7cm}
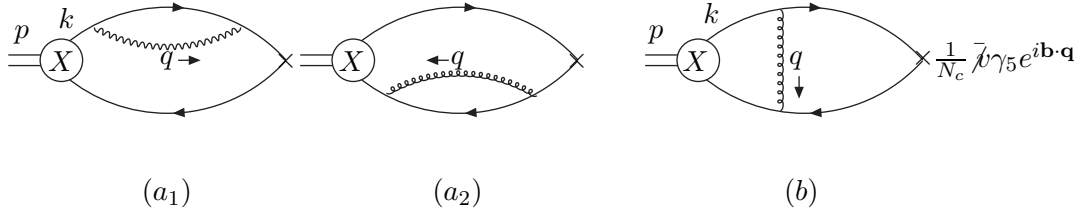
{\bf Fig.2} One loop diagrams for the meson wave function ${\cal P}$. 

\bigskip

Contribution to ${\cal P}$ from vertex correction, Fig.~2 graph($b$): \\
The vertex correction to the meson wave function is  
\begin{equation}
{\cal P}_{(b)} = i (4 \pi \alpha_s) C_F \, \int \frac{ d^4 q}{(2 \pi)^4} \, 
e^{i \b \cdot \q} \; {\rm tr} \left[ {\bf X} \gamma^\nu 
\frac{1}{-\not{p}+\not{k}-\not{q}+i \epsilon} \bar{\not{v}} \gamma_5 
\frac{1}{\not{k}-\not{q}+ i \epsilon} \gamma^\mu \right] 
N_{\mu \nu} \frac{1}{q^2-\lambda^2+ i \epsilon}  \, . 
\label{Pbexpr}
\end{equation}
The impact parameter $b$ acts as an ultraviolet regulator 
so the loop integral can be kept at $D=4$ dimensions.  
Standard manipulations for the $-g_{\mu \nu}$ piece of the gluon propagator 
yield
\begin{equation}
{\cal P}_{(b)}|_{-g_{\mu \nu}} = -\frac{\alpha_s}{4 \pi} C_F {\cal P}^{(0)}
\left[\ln(b^2 \lambda^2) +\frac{1}{6} b^2 \lambda^2 \ln(b^2 \lambda^2)  
\right]
+{\cal O}(b^4 \lambda^4 \ln(b \lambda))+{\cal O}( \ln^0(b \lambda))  \, .
\label{feynresult}
\end{equation}
Without the gluon mass regulator, the $\ln(b^2 \lambda^2)$ term would 
turn into an IR divergence that should be dimensionally regularized. 
Here, as well as in the rest of the calculations the interest lies in 
the presence of the $b^2\lambda^2 \ln(b^2 \lambda^2)$ nonanalytic term. 

The calculation involving the $n$-dependent pieces of the gluon propagator
is somewhat subtler. 
Evaluation of the traces in Eq.~(\ref{Pbexpr}) yields 
\begin{equation}
{\cal P}_{(b)}|_{n_{(\mu} q_{\nu)}} = i (4 \pi \alpha_s) C_F {\cal P}^{(0)} \, 
(I_1+I_2) \, , 
\label{vax1}
\end{equation}
with
\begin{equation}
I_1 = \int \frac{d^4 q}{(2 \pi)^4} e^{i \b \cdot \q} \, 
\frac{ [n\cdot (q-k) + v\cdot n \, \bar{v} \cdot (q-k) - 
\bar{v} \cdot n \, v \cdot q] \, n \cdot q}
{[(q-k)^2+i \epsilon] \,  (q^2-\lambda^2 +i \epsilon) 
\,[(n \cdot q)^2 +\eta^2]}
\label{I1def}
\end{equation}
and 
\begin{equation}
I_{2} = I_{1}|_{k \rightarrow k-p} \, .
\label{I2def}
\end{equation}
The two integrals $I_1$ and $I_2$ arise from the two pieces 
$n_\mu q_\nu$ and $n_\nu q_\mu$ of the gluon propagator respectively. 
At this stage it is advantageous to {\it fix the gauge} by the choice
\begin{equation}
n^\mu = \frac{1}{\sqrt{2}} \left( v^\mu -\bar{v}^\mu \right) \, .
\label{gaugefix}
\end{equation}
This gauge choice has the advantage of simplifying the corresponding integrals 
by removing all dependence on $q^-$ and $\q$, 
from the numerator of Eq.~(\ref{I1def}),  i.e. the normal coordinates in the 
collinear limit for the gluon momentum.
In this gauge $I_1$ becomes
\begin{equation}
I_1= \int \frac{d^4 q}{(2 \pi)^4} e^{i \b \cdot \q} \, 
\frac{ [ 2 n\cdot q + (k \cdot q)/(n \cdot k) -2 n \cdot k] \, n \cdot q}
{[(q-k)^2+i \epsilon] \, (q^2-\lambda^2+i \epsilon) 
\,[(n \cdot q)^2 +\eta^2]} \, .
\label{I1fix}
\end{equation}
The above integral is evaluated via the introduction of Schwinger 
parameters. The intermediate steps can be reconstructed by using the  
techniques and the integral relations presented in Ref.~\cite{CapLeib}
and the result is
\begin{equation}
 I_1 = \frac{i}{(4 \pi)^2} \, 
\int_0^1 d x_1 \, x_1^{-1/2} \int_0^1 d x_2  
\left( K_0(\sqrt{B}) + A  \frac{1}{\sqrt{B}} K_1(\sqrt{B}) \right) \, ,
\label{axial8}
\end{equation}
with
\begin{eqnarray}
A &=& x_2  (-2+x_2+x_1 x_2) b^2 (n \cdot k)^2  \, ,
\nonumber \\
B &=& b^2 \left[ (1-x_1) x_2^{\,2} (n \cdot k)^2 + (1-x_2) \lambda^2 + 
\frac{1-x_1}{x_1} \eta^2 \right] \, .
\label{Btrans} 
\end{eqnarray}
Since we are interested in the limit $\lambda \rightarrow 0$ 
we will calculate leading and nonanalytic $\lambda$-dependence of the 
Feynman integrals with the help of the Mellin transformation defined as  
\begin{equation} 
M[F(t)](\beta) = \int_0^\infty d t \, t^{-\beta-1} F(t) \, .
\label{Mellindef}
\end{equation}
It should be noted that the Mellin transform method has been used extensively 
in the study of the high energy behavior of Feynman diagrams. 
For a review of the method and for further references, see \cite{ELOP}. 
The definition~(\ref{Mellindef}) yields the following correspondence between 
poles in $\beta$ and (large) $t$ dependence: 
\begin{equation}
M[F](\beta) = \frac{r}{(\beta- \beta_0)^n} \leftrightarrow 
F(t) = \frac{r t^{\beta_0} \ln^{n-1}t}{\Gamma(n)} \, .
\label{correspondence}
\end{equation}
Nonanalytic terms in $t$ (logarithmic) are generated by at least double poles 
in the Mellin image. We define the following
dimensionless ratios 
\begin{equation} 
t = \frac{(n \cdot k)^2}{\lambda^2} \, , \  \  \  \  \
s = b^2 (n \cdot k)^2 \, , 
\label{tsdef} 
\end{equation}
and we note that the parameter $\beta$ also acts as an IR regulator. 
Then, in Eqs.~(\ref{Btrans}) the $\eta \rightarrow 0$ limit for the 
gluon propagator can be taken. 
The Mellin transformation of the Bessel $K$ functions can be found in  
Ref.~\cite{GR} and the Mellin image of Eq.~(\ref{axial8}) turns out to be 
\begin{equation}
M[I_1] = \frac{i}{(4 \pi)^2} \left(J_0 + J_1 \right) \, ,
\label{IJJ}
\end{equation}
where
\begin{eqnarray}
J_0(\beta) &=&   2^{\beta} \Gamma(\beta) s^{-\beta/2} 
\int_0^1 d x_1 \, x_1^{\beta/2} (1-x_1)^{-1/2} 
\int_0^1 d x_2 \, x_2^{\beta} (1-x_2)^{-\beta} 
K_{-\beta} (\sqrt{s x_1} x_2)  \, ,
 \\ 
J_1(\beta) &=& 2^{\beta} \Gamma(\beta) s^{1/2-\beta/2} 
\int_0^1 d x_1 \, x_1^{-1/2+\beta/2} (1-x_1)^{-1/2}
\nonumber \\ 
&\quad& \times \int_0^1 d x_2 \, x_2^{\beta} (1-x_2)^{-\beta} 
(-2+2x_2-x_1 x_2) \,  K_{1-\beta} (\sqrt{s x_1} x_2)  \, .
\label{mellin3}
\end{eqnarray}
Inspecting the above integrals we see that for $\beta > 0$ there are no 
singularities. Singularities arise only for $\beta \le 0$ and they are 
generated from the integration end points $x_1 \rightarrow 0$ and/or 
$x_2 \rightarrow 0$.
The extraction of such singularities is straightforward because the series 
expansions of the $K$-functions for small argument can be used.  
From Eq.~(\ref{correspondence}) it follows that the relevant nonanalytic 
terms that we are interested in can arise from at least double poles at 
$\beta=0, -1$ \footnote{ It is straightforward to see that there are no other 
points in the range between -1 and 0 that would give double poles, thus 
excluding any linear in $\lambda$ corrections arising from poles at  
$\beta=-1/2$.}. 
We examine these two points in turn. 

For the poles at $\beta=0$ we set $\beta=\delta$, $\delta \rightarrow 0$ 
and obtain
\begin{eqnarray}
J_0(\beta=\delta) &=& {\cal O}(\frac{1}{\delta}) \, , 
\nonumber \\
J_1(\beta=\delta) &=& -\frac{1}{\delta^3} 
+\frac{2}{\delta^2}(1-\ln 2)+{\cal O}(\frac{1}{\delta}) \, .
\label{J10final}
\end{eqnarray}
Combining Eqs.~(\ref{J10final}, \ref{IJJ}) and inverting 
the Mellin transformation we get 
\begin{equation}
I_1|_{\beta \rightarrow 0} = \frac{-i}{(4 \pi)^2} \left[ 
\frac{1}{2} \ln^2 \frac{\lambda^2}{(n \cdot k)^2} 
+ 2(1-\ln 2) \ln \frac{\lambda^2}{(n \cdot k)^2} \right]
+ R \, .
\label{I10}
\end{equation}
The residue $R$ generically contains all terms that are analytic in 
$\lambda^2$.
Similarly, the poles are $\beta=-1$ are obtained by setting 
$\beta=-1+\delta$ and power expanding the $K$-functions. 
The results are 
\begin{equation}
J_0(\beta=-1+\delta) = (s{\rm -independent}) \, , 
\label{J01inter}
\end{equation} 
and
\begin{equation}
 J_1(\beta=-1+\delta)  =   
(s {\rm  -independent}) -\frac{s}{4 \delta^3} 
+\frac{s}{2 \delta^2} (1 -\ln2) +{\cal O}(1/\delta) \, .
\label{J11inter}
\end{equation}
Again, combining the above two equations with Eq.~(\ref{IJJ}) and inverting 
the Mellin transformation we have
\begin{eqnarray}
I_1|_{\beta \rightarrow -1} &=& 
(b{\rm -independent})
\nonumber \\
&\quad& +\frac{i}{(4 \pi)^2} \left[ -\frac{1}{8} b^2 \lambda^2 
\ln^2 \frac{\lambda^2}{(n \cdot k)^2} 
- \frac{1}{2}(1-\ln2)  b^2 \lambda^2 
\ln \frac{\lambda^2}{(n \cdot k)^2}  \right] + R
\nonumber \\
&=& (b{\rm -independent})
\nonumber \\
&\quad -& \frac{i}{(4 \pi)^2} b^2 \lambda^2 \left[ \frac{1}{8}
\ln^2 \left( \frac{\lambda^2}{ x^2 Q^2 } \right)
+\frac{1}{2}(1- \ln 2) \ln \left( \frac{\lambda^2}{ x^2 Q^2}\right) \right] 
+R \, .
\label{I11}
\end{eqnarray}
In the last step we have used that $(n \cdot k)^2=x^2 \, Q^2$.
By ($b$-independent) we denote all terms that are non-analytic in $\lambda^2$, 
${\cal O}(\lambda^2 \ln \lambda^2)$ in this case, and are independent of $b$. 
Analytic in $\lambda$ terms reside in $R$. We make this distinction because, 
as we will see later, the terms denoted by ($b$-independent) are of 
infrared origin and cancel against the self energy contribution. 
The expression for $I_2$, Eq.~(\ref{I2def}), is obtained 
from the above by simply $x^2 \rightarrow (1-x)^2$. 
The overall contribution to the vertex from the $n_{(\mu} q_{\nu)}$ 
piece of the gluon propagator is, Eq.~(\ref{vax1}),
\begin{eqnarray}
{\cal P}_{(b)}|_{n_{(\mu} q_{\nu)}} &=& \frac{\alpha_s}{4 \pi} \, 
C_F \, {\cal P}^{(0)} \,  b^2 \lambda^2 
\left[ \frac{1}{8} \left( \ln^2 \frac{\lambda^2}{x^2 Q^2} 
+\ln^2 \frac{\lambda^2}{(1-x)^2 Q^2} \right) 
+ (1- \ln 2) \ln  \frac{\lambda^2}{ x (1-x) Q^2}
\right] 
\nonumber \\
&\quad& +( b-{\rm independent}) +R  \, .
\label{vertexaxial}
\end{eqnarray}
Finally, the $q_\mu q_\nu$ piece of the gluon propagator
gives $b$-independent leading contributions and hence it cancels against 
the corresponding piece of the self energy, as we will point out below.  

Contribution to ${\cal P}$ from self-energy correction, 
Fig.~2 graphs $(a_1)$ and $(a_2)$: \\
Due to UV divergences we employ dimensional regularization with 
$D=4-2 \epsilon$ and renormalization scale $\mu$.
\begin{equation}
{\cal P}_{(a_1)} = i (4 \pi \alpha_s) C_F \, \mu^{2 \epsilon}
\int \frac{ d^D q}{(2 \pi)^D} \, 
{\rm tr} \left[ {\bf X} \bar{\not{v}} \gamma_5 
\frac{1}{\not{k}+i \epsilon} \gamma^\nu 
\frac{1}{\not{k}-\not{q}+i \epsilon} \gamma^\mu \right] 
N_{\mu \nu} \frac{1}{q^2-\lambda^2+i \epsilon} 
\label{P1expression}
\end{equation}
It is straightforward to find that, even before any gauge fixing, 
the self energy contribution is such that
\begin{equation}
{\cal P}_{(a_1)} + {\cal P}_{(a_2)}
 + {\cal P}_{(b)}(b \rightarrow 0)  = R_{UV} \, ,
\label{seresult}
\end{equation}
where the residue $R_{UV}$ contains terms of purely ultraviolet origin and it 
is due to the mismatch in the UV regularization of the two sets of diagrams. 
Note that the vertex diagram is UV regularized by a finite $b$ whereas the 
self energy diagrams are dimensionally regularized.
Had we taken the limit $b \rightarrow 0$ before calculating the vertex 
integrals, then this would have required to also dimensionally regularize 
the vertex and then in Eq.~(\ref{seresult}) $R_{UV}$ would have been exactly
zero. In any case, the net result of this analysis is that the self 
energy diagrams will cancel all $b$-independent terms in the vertex, such as 
the ones shown in Eq.~(\ref{vertexaxial}) as well as the vertex contributions 
from the $q_\mu q_\nu$ piece of the gluon propagator. 
 
The final result for the one loop radiative correction to the wave function 
in the massive gluon scheme is obtained by combining the partial results 
of Eqs.~(\ref{feynresult}, \ref{vertexaxial}) and (\ref{seresult}).
Retaining only the nonanalytic in $\lambda$ terms up to ${\cal O}(\lambda^2)$  
we obtain:
\begin{equation}
{\cal P}^{(1)} =  \frac{\alpha_s}{4 \pi} \, C_F \, {\cal P}^{(0)} \,  
\left( C_1 b^2 \lambda^2 \ln \lambda^2 + C_2 b^2 \lambda^2 \ln^2 \lambda^2 
\right) + R  \, ,
\label{final}
\end{equation}
with 
\begin{equation} 
C_1 = -\frac{1}{2} \ln (x (1-x) Q^2)  \, , 
\ \ \ \ C_2=\frac{1}{4} \, .
\label{coeffs}
\end{equation}
The familiar Sudakov factor $S_{wf} \sim \alpha_s \ln^2 Q^2$, 
calculated for fixed coupling $\alpha_s$, is analytic in $\lambda^2$ and it
is contained in the residue $R$.
In the coefficients  $C_1, C_2$ we have only kept the parts that are 
$n$-independent to leading logarithmic in $Q$ order. 
Note that the $n$-independent leading contributions to $C_1$, $C_2$ come 
only from Eq.~(\ref{vertexaxial}). 
The $n$-dependence enters through the combination 
\mbox{$\ln(n \cdot k)^2 + \ln(n \cdot (k-p))^2$}, and any variation of $n$ 
will lead to change that is subleading in $Q$, i.e. it will be down by a 
$\ln (x(1-x)Q^2)$.   
Fixing the gauge as in Eq.~(\ref{gaugefix}) organizes conveniently 
the leading and subleading $\ln Q^2$ contributions but the result that 
we quote in Eqs.~(\ref{final}, \ref{coeffs}) is the gauge independent 
leading piece. 

We now return to the question of regularizing infrared singularities in the 
axial gauge and within the massive gluon scheme. 
In the massive gluon propagator, Eq.~(\ref{propdef}), we used the parameter 
$\lambda$ to regularize on-shell singularities whereas the polarization tensor 
$N_{\mu \nu}$ was defined through the principal value prescription and the 
regulator $\eta$. 
If we treat the parameter $\lambda$ as a lagrangian mass, instead of a 
dispersive parameter (this is at the one loop level only) then the propagator 
will be modified by the addition of a fourth term of the form
$ n_\mu n_\nu \lambda^2 /[(n \cdot k)^2-\lambda^2 n^2]$.
A similar prescription has been considered in the leading 
$\lambda \rightarrow 0$ limit in Ref.~\cite{Landshoff}. 
However, beyond the logarithmic in $\lambda$ level,
this term yields $b$-dependent ${\cal O}(\lambda^2 \, \ln^3 \lambda^2)$ 
contributions that cannot be interpreted as arising from some pinch 
singularity in the limit $\lambda \rightarrow 0$. 
Such artifacts of the regularization prescription are avoided once one keeps 
in mind that $\lambda$ is not a lagrangian gluon mass but just an on-shell 
regulator, as it appears in Eq.~(\ref{propdef}).

\subsection{The running $\alpha_s$ scheme and IR renormalons}

The running $\alpha_s$ scheme for estimating nonperturbative corrections 
starts with the expression for the one loop radiative corrections to the 
meson wave function. The leading logarithmic term is well known and 
can be readily obtained from the expressions in Eqs.~(\ref{Pbexpr}, 
\ref{P1expression}) after rationalizing the fermion propagators and 
applying the collinear approximation to the numerator factors. 
Equivalenlty it can be computed by calculating emission from two, 
almost parallel, Wilson lines. 
All infrared divergences are dimensionally regularized and no $\lambda$ 
regulator need be introduced. Principal value prescription is used for 
the axial gauge gluon propagator which now reads 
\begin{equation}
D_{\mu \nu}(q, n) = \frac{i}{q^2+i \epsilon}  \,
\left[ - g_{\mu \nu} +\frac{1}{n \cdot q} \, 
n_{\left( \mu \right. } q_{\left. \nu \right)} 
- \frac{n^2}{(n \cdot q)^2} \, q_\mu q_\nu \right] \, .
\label{propdefmassless}
\end{equation}
After performing the collinear approximation to the numerator factors 
and then integrating over $q^-$ by closing around its poles in the 
$q^-$ complex plane we obtain the result   \cite{BottsSter}
\begin{equation}
{\cal P}^{(1)}(x, b, p,\mu) = \frac{\alpha_s}{2 \pi^2} C_F 
{\cal P}^{(0)}(x, b=0, p,\mu)
\int^{\q^2=Q^2} \frac{d^2 \q}{\q^2}
\left(e^{i\b \cdot \q}-1 \right) 
\int_{|\q|}^{x p^+} \frac{dq^+}{q^+} 
+ (x \rightarrow 1-x) \, .
\label{finP}
\end{equation}
The Sudakov exponent due to perturbative evolution of the wave function, 
Eq.~(\ref{exponentiate}), is determined to leading order in $\alpha_s$ 
by the above expression. 
At this stage the running coupling is introduced. A perturbative analysis
at the logarithmic level \cite{BottsSter} indicates that 
the scale of the coupling is set by the
transverse momentum of the emitted  gluon and it encodes the information that
the strength of the interactions  increases for emission at large impact
parameter.  The Sudakov exponent now reads
\begin{eqnarray}
S_{wf} &=& -\frac{C_F}{2 \pi^2} \int \frac{d^2 \q}{\q^2} \, \alpha_s(\q^2) \,
\left( e^{i \b \cdot \q}-1 \right) \,
\left( \int_{|\q|}^{x p^+} \frac{dq^+}{q^+}
+\int_{|\q|}^{(1-x) p^+} \frac{dq^+}{q^+} \right)
\nonumber \\
&=& -\frac{C_F}{2 \pi^2} \int \frac{d^2 \q}{\q^2} \alpha_s(\q^2) \,
\ln \frac{x (1-x) (p^+)^2}{\q^2} \,
\left( e^{i \b \cdot \q}-1 \right) \, .
\label{sudexp}
\end{eqnarray} 
When the lower limit of the $d \q^2$ integration is set to $1/b^2$ the result 
in Eq.~(\ref{sexplog}) is recovered. 
However, apart from the case of small size onia, for the usual light mesons  
$1/b \sim \Lambda_{\rm QCD}$.   
It is then apparent that after the introduction of the running coupling the 
Sudakov exponent will become dominated by the lower end of the $d \q^2$ 
integration where perturbation theory itself is ill defined. 
Nevertheless, what we are interested in here is not the numerical stability 
of the perturbative results but signals of nonperturbative corrections. 
We therefore proceed by allowing the infrared regulator to become 
${\cal O}(\Lambda_{\rm QCD})$ and  introduce the Borel transformation of the 
Sudakov exponent defined as 
\begin{equation}
S_{wf}(x, b, Q,\mu; \alpha_s) =
\int_0^\infty d \sigma \, \tilde{S}_{wf}(x, b, Q, \mu; \sigma) \,
e^{-\sigma/\alpha_s} \, .
\label{boreldef}
\end{equation}
The Borel image $\tilde{S}_{wf}$ can be obtained by first using the following
representation of the  one loop running coupling:
\begin{equation}
\alpha_s(\q^2)=\int_0^\infty d \sigma \, e^{-\sigma \beta_0 
\ln(\q^2/\Lambda^2)} \, .
\label{boralpha}
\end{equation} 
Substituting the above expression into Eq.~(\ref{sudexp}) 
and integrating over $\q$ we get
\begin{equation}
S_{wf} =\frac{C_F}{2 \pi^2} \,
\ln \frac{x (1-x) (p^+)^2}{\Lambda^2} \, \int_0^\infty d \sigma \,
\frac{\Gamma(1-\sigma \beta_0 )}{\sigma \beta_0  \,
\Gamma(1+\sigma \beta_0)} \, e^{-\sigma/\alpha_s(4/b^2)} \, .
\label{expalpha}
\end{equation}
Comparing the result with the definition (\ref{boreldef}) we read off 
the Borel image 
\begin{equation}
\tilde{S}_{wf} =  \frac{C_F}{2 \pi^2} \,
\ln \frac{x (1-x) (p^+)^2}{\Lambda^2} \, 
\frac{\Gamma(1-\sigma \beta_0)}{\sigma \beta_0 \,
\Gamma(1+\sigma \beta_0)} \, .
\label{borimage}
\end{equation}
The pole at $\sigma=0$ is of UV origin. 
Infrared sensitivity is parametrized by the presence of IR renormalon poles at 
$\sigma=n/\beta_0 \, , \ n=1,2,...$. 
The leading IR renormalon is at $\sigma=1/\beta_0$ which leads to 
${\cal O}(\Lambda_{\rm QCD}^2 b^2)$ power corrections of the form
\begin{equation}
S_{wf}= S_{wf}^{PT} + C \Lambda_{\rm QCD}^2 b^2 \, 
\ln \frac{x (1-x) Q^2}{\Lambda_{\rm QCD}^2} \, .
\label{nonpert}
\end{equation}
Terms that are $Q$-independent have been omitted in the above equation.
As in the case of the massive gluon scheme, it is this part of the coefficient 
that is $n$-independent to leading logarithmic in $Q$ order and it arises 
from soft gluon emission, hence it is universal. 
The coefficient $C$ contains the principal value integral over $\sigma$, 
which is one possible way to define it. This ambiguity 
is anticipated since perturbation theory is ill defined and
therefore  it cannot predict the size of nonperturbative corrections, only
their  scaling with $Q$. We emphasize that the manipulations that
led to  Eq.~(\ref{borimage}) are formal and the transformation cannot be
inverted back  to $\alpha_s$ space without the use of some prescription (or
convention).  The value of the coefficient $C$ will depend on this 
prescription. 
In addition, as with the massive gluon case, higher orders in $\alpha_s$ are
not suppressed, unless one invokes an additional assumption of say the freezing
of the coupling at small momenta \cite{dispersive, freeze}.

The connection between the running coupling and the massive gluon
calculation is established by comparing  Eq.~(\ref{nonpert}) with
Eq.~(\ref{final}).  Note that in the massive gluon result we have not written
the analogue of the 
$S^{PT}_{wf}$ that we see above. 
This is because in the Mellin transformation analysis of the massive gluon 
integrals we looked only for contributions that are nonanalytic in the
regulator $\lambda^2$ and not in the impact parameter $b^2$. 
Had we chosen the latter we would have generated the perturbative 
answer $S^{PT}_{wf}$ in the massive gluon scheme as well. 
The nonperturbative pieces in the two calculations can be mapped onto 
each other, up to a multiplicative constant, via the identification
\begin{equation} 
\lambda^2 \ln \lambda^2  \leftrightarrow 
\Lambda_{\rm QCD}^2  \, . 
\label{map1}
\end{equation}
In Eq.~(\ref{final}), the $\lambda^2 \ln ^2 \lambda^2$ term indicates that the 
coefficient of the term $\lambda^2 \ln \lambda^2$ will depend on $\ln Q^2$, 
$Q^2$ being the only other scale involved, hence we also identify 
\mbox{$ (1/2) \lambda^2 \ln^2 \lambda^2  \leftrightarrow 
\Lambda_{\rm QCD}^2 \ln \Lambda_{\rm QCD}^2$}. 
The existence of $\lambda^2 \ln \lambda^2$ contributions is a signal for the 
presence of a condensate-like term. Had only this term been present then one 
could expect that the power corections could be captured from an O.P.E. 
in terms of local operators. 
The existence of $\lambda^2 \ln^2 \lambda^2$ contributions, though, signals 
that such an expansion is not possible for the wave function of a light 
hadron. An operator expansion formalism for the wave function must 
necessarily involve nonlocal operators that extend along the light cone 
and, in analogy with the usual OPE, their expectations are to be parametrized 
by nonlocal condensates.
The possibility of parametrizing power corrections to the meson 
electromagnetic form factor in terms of nonlocal condensates 
has been discussed in Ref.~\cite{BakRad}. 

The other important piece of information in the leading power correction term 
is that its $b^2$ dependence leads to a Gaussian factor for the meson wave 
function. This is obtained from Eqs.~(\ref{exponentiate}) and (\ref{nonpert}).
\begin{equation}
{\cal P} = {\cal P}^{PT} \,
\exp \left( -C \ln \left( \frac{x (1-x) Q^2}{\Lambda_{\rm QCD}^2} \right)
 \, \Lambda_{\rm QCD}^2 b^2 \right) \, ,
\label{gaussian}
\end{equation}
with ${\cal P}^{PT}$ denoting the wave function perturbatively evolved with 
exponent given in Eq.~(\ref{sexplog}). 
We turn to a discussion of this form in the next section.

\section{The Gaussian form of the meson wave function} 
\setcounter{equation}{0}

The wave function in Eq.~(\ref{final}) and Eq.~(\ref{gaussian}) can be 
rewritten in the following suggestive form:
\begin{equation} 
{\cal P}(x, b, Q) = \phi(x, Q) \exp \left( -S_{wf}^{PT} - 
b^2 S_2(x (1-x)Q^2) - b^4 S_4(x (1-x)Q^2) +... \right) \, .
\label{expform}
\end{equation} 
Infrared renormalons at integer values generate a $b^2$ expansion 
for the Sudakov exponent with $Q$ dependent coefficients. 
This form is the same as the one seen in the Drell-Yan process at measured 
$Q_\perp$ and has been derived by Korchemsky and Sterman in 
Ref.~\cite{KorchSter}. 
These authors have also given an operator definition of the power corrections
in terms of pairs of Wilson lines and their transverse derivatives. 
For the same Drell-Yan process, the Gaussian form in impact parameter 
space has been noted years ago and outside the renormalon context 
by Collins and Soper \cite{CollinsSoper} in their treatment of the 
infrared sensitivity of the Sudakov exponent. 
 
The dimensionfull functions $S_2$, $S_4$, ... cannot be normalized 
within perturbation theory. 
However, we have obtained some information beyond the summation of 
logarithmic corrections. 
If we assume that in Eq.~(\ref{gaussian}) the unknown coefficient $C$ 
is a number of ${\cal O}(1)$ and that our methods capture correctly 
the $x$-dependence of the leading nonpeturbative corrections then 
\begin{equation}
S_2(Q) \sim \Lambda^2_{\rm QCD} \ln(x (1-x) Q^2) \, ,
\label{s2}
\end{equation}  
where only the universal piece is retained, as it has been argued in 
the previous section.  
For fixed $x \ne 0$, power corrections lead to increasing suppression 
of emission at large impact parameter with increasing $Q$. 
This suppression is {\it in addition} to the one generated by  
$S_{wf}^{PT}$. 
However, for fixed $Q$, it is seen that this additional suppression 
becomes weak in the end-point region. 
The end-point region is enhanced relative to the 
central region by the Gaussian dependence of the meson wave function. 
This is not surprising since it is in the end-point region that
the effective hard scale $x (1-x)Q^2$ becomes small and 
power corrections become more important. 
We emphasize that all this is on top of the Sudakov suppression,
$S_{wf}^{PT}$, as mentioned above. 
In this respect, the pattern of power corrections is very similar to that 
observed, for example in event shape variables where the leading power 
corrections \cite{eventshape} also come from the end-point region
(two jet limit in this case), which is itself Sudakov suppressed.

For exclusive processes there exist in the literature various models for the
wave function with which we may compare our results. 
In the case of the meson, the most popular is the one obtained  from 
an oscillator model of two constituent quarks  boosted in the light cone 
frame. This leads to a wave function of the  form \cite{BHLM}, 
\cite{Zhitnisky}
\begin{equation}
{\cal P}(x, b, Q) = \phi(x, Q) \exp \left( -S_{wf}^{PT} - 
<k_\perp^2> x (1-x) \, b^2 \right)  \, , 
\label{oscmod} 
\end{equation}
where the average $ <k_\perp^2>^{-1}$ is the oscillator parameter. 
Note that this Gaussian form is such that it can interpolate between the 
perturbative tail of the wave function given by $\exp(-S_{wf}^{PT})$ and 
the nonperturbative region. The ${\cal O}(b^2)$ term in the exponent has been 
found numerically important  in order for Sudakov resummed perturbative 
expressions for the meson form factor to be applicable in the subasymptotic
$Q$ region \cite{JakobKroll}.  
It is therefore interesting to compare its $x$-dependence with 
the expression we get from perturbation theory. 
We observe that like in Eq.~(\ref{s2}), the exponent in Eq.~(\ref{oscmod}) 
enhances the end-point regions compared to the central region. 
However, from the renormalon based approach, which is rooted in perturbation
theory, one can expect a  dependence on $x(1-x)$ which is logarithmic but
not proportional to it.  
The analysis of nonperturbative corrections as signaled by perturbation theory 
and phenomenological models for the meson wave function in the
nonperturbative regime have similar qualitative behavior although the 
end-point enhancement in the former case is much milder than in the latter.
For our approach to give results in exact correspondence with, 
say, the oscillator model, 
it would require a resummation of the higher order in $\alpha_s$ 
contributions to the $\lambda^2 \ln \lambda^2$ term at the  level of the 
{\it exponent}.
These  could turn the $\ln x(1-x)$ into a $x^n(1-x)^n$ dependence.
It is not obvious to us how such a resummation would be implemented.  
Recall that in the region where nonperturbative corrections are  generated, 
the coupling is  $\alpha_s={\cal O}(1)$ and there appears to be no small 
parameter around which to build and then resum a perturbative series.  
More importantly, it is not known how to define the massive gluon scheme 
for higher loop corrections, and for the running coupling scheme it is only an 
assumption that $\alpha_s=\alpha_s(k_\perp^2)$ all the way into the deep 
infrared region.
We can say for sure, though, that any answer obtained from perturbation theory
would necessarily have $Q$ dependence in the Gaussian form that it generates,
 unlike Eq.~(\ref{oscmod}) which is $Q$-independent beyond the logarithmic
corrections residing in $S_{wf}^{PT}$.
It should be pointed out here that, recently, renormalon based models have 
been used to predict the $x$-dependence of the higher twist structure 
functions in deeply inelastic  scattering with some phenomenological success 
\cite{dis} and this was one of our motivations for pursuing the above analysis.

\section{Summary}

In this paper we studied the nonperturbative corrections to the meson 
wave function using the methods that have already been developed for 
semi-inclusive cases, namely the massive gluon and the renormalon methods. 
We found that the leading nonperturbative corrections are of order 
${\cal O}(\Lambda_{\rm QCD}^2 b^2)$ for the two-quark wave function 
at transverse separation $b$. 
The exponentiation of such contributions leads to a Gaussian factor 
in addition to the Sudakov resummed logarithmic enhancements. 
Of particular interest is the $x$-dependence of this Gaussian factor.
It leads to the conclusion that the power corrections arise from the end-point 
regions. 
This $x$-dependence has been compared with low energy Gaussian models for 
the wave function.

It must be emphasized that both methods for obtaining the leading 
nonperturbative corrections  have their origin in perturbation theory. 
They predict correctly the type of the power correction but since they are 
applied in a region where the coupling is normalized at low scales 
they have limited predictability for the coefficient of the power corrections 
unless additional assumptions are introduced, such as freezing of the 
coupling.
Thus our predictions for the $x$-dependence should be considered as another 
model. 
Similar approach for the $x$-dependence of the higher twist structure 
functions in D.I.S. has met with some phenomenological success.
It would be interesting to apply this model to the phenomenological study 
of the meson electromagnetic form factor and elastic scattering.

\medskip

{\it Acknowledgements:} 
We would like to thank Profs. P. Kroll and N. G. Stefanis for a communication. 
This work was supported in part by the US Department 
of Energy. One of us (AS) gratefully ackowledges a fellowship from the 
Alfred P. Sloan Foundation.

\end{document}